\documentstyle[prl,aps,epsf,twocolumn]{revtex}

\tighten

\begin{document}
\draft
\twocolumn[\hsize\textwidth\columnwidth\hsize\csname
@twocolumnfalse\endcsname

\title{\bf Dynamics of Lennard-Jones clusters: A characterization of the
activation-relaxation technique}

\author{Rachid Malek\protect\cite{malekadd} and Normand
Mousseau\protect\cite{moussadd}}

\address{
Department of Physics and Astronomy and CMSS, Ohio University, Athens,
OH 45701, USA }

\date{\today}

\maketitle

\begin{abstract}

The potential energy surface (PES) of Lennard-Jones clusters is
investigated using the activation-relaxation technique (ART).  This
method defines events in the configurational energy landscape as a
two-step process: (a) a configuration is first activated from a local
minimum to a nearby saddle-point and (b) is then relaxed to a new
minimum.  Although ART has been applied with success to a wide range
of materials such as {\it a}--Si, {\it a}-SiO$_2$ and binary
Lennard-Jones glasses, questions remain regarding the biases of the
technique.  We address some of these questions in a detailed study of
ART-generated events in Lennard-Jones (LJ) clusters, a system for
which much is already known.  In particular, we study the distribution
of saddle-points, the pathways between configurations, and the
reversibility of paths.  We find that ART can identify all
trajectories with a first-order saddle point leaving a given minimum,
is fully reversible, and samples events following the Boltzmann weight
at the saddle point.

\end{abstract}

\vskip2pc
]

\vspace*{-0.5cm}
\narrowtext

\section{Introduction}

In many materials and systems, microscopic structural relaxation takes
place on time scales much longer than those of the atomistic
oscillations set by the phonon vibrations (10 ps).  This is the case,
for example, for glasses and other complex materials.  Such a time
spread in the dynamics can be understood from the configurational
energy landscape picture.  Indeed, the system finds itself in a deep
minimum surrounded by energy barriers much higher than its thermal
energy and only rare fluctuations can allow the system to jump over a
barrier and move to a new minimum.  These long time scales are
especially prohibitive for numerical studies using traditional methods
such as molecular dynamics (MD) and real space Monte Carlo, which are
tied to the phonon time scale.

One way to reach this long time scale is through activated
dynamics ~\cite{Voter,BH,BM1}. In this case, the algorithm focuses
directly on the appropriate mechanisms and describes the dynamics
as a sequence of metastable states separated by energy barriers.
These metastable configurations can be well identified by their
atomic positions at zero K, which correspond to a local minimum
in the configurational energy landscape.  Knowledge of the
distribution and properties of these local minima is sufficient
to determine the thermodynamical properties of the system. A proper
description of the dynamics, however, also requires knowledge of the
rates controlling jumps from one local minimum to another.

Activated dynamics has been successfully applied to a range of
discrete problems.  It has turned out to be especially useful in the
study of metallic surfaces, where it is possible to identify and
compute {\it a priori} the whole set of barrier to be visited during
the dynamics ~\cite{KMC}.  These methods have also be applied, with
further approximations, to other systems such as the hetero-epitaxial
growth of semiconductor compounds ~\cite{djaf1}.  Such
simulations can reach simulated times 10 or 12 orders of magnitude
longer than what can be done with molecular dynamics.

For more complex systems, however, identifying the barriers and insuring
proper statistical sampling remain a challenge.   The
activation-relaxation technique (ART)~\cite{BM1,MB2} was proposed
recently to address this challenge. A number of questions remain
regarding the biases of the method,  how it samples the potential energy
landscape and whether or not it is reversible.

Before achieving the long term objective of developing an
algorithm for simulating the atomistic time evolution of complex
materials, it is necessary to address these issues.  Here, we
answer some of these questions in a study of Lennard-Jones (LJ)
clusters, comparing ART of Barkema and Mousseau with a similar
algorithm introduced by Doye and Wales(DW)~\cite{DW}.  In
particular, we look at the sampling of barriers, the
reversibility of the paths from the saddle-point and from the new
minimum. The dynamical and thermodynamical properties of LJ
clusters have been thoroughly studied ~\cite{pt1,pt2,pt3} and they provide
an ideal model for the development of global optimization
techniques~\cite{lj1,lj2,lj3} as well as activated methods.

This Paper is constructed as follows. In the next section, we
describe the details of the activation-relaxation technique. We
then present the results of our simulation on 3 different LJ
clusters (13, 38 and 55 atoms) and give a short discussion.

\section{Technique}

The activation-relaxation technique (ART),~\cite{BM1,MB2} is a generic
method to explore the surface energy landscape and search for
saddle-points.  It has been applied with success to amorphous
semiconductors, silica and metallic glasses
~\cite{BM1,MB2,MB3,MBL,ML,MB4}.

ART defines events in the configurational energy landscape
according to a two-step process: (a) starting from a local
minimum, a configuration is pushed up to a local saddle point,
representing the activation process; (b) from this saddle point,
the configuration is relaxed into a new minimum; this whole
process is called an event.  For the activation, we use a
modified version of the algorithm, ART {\it nouveau }, which was
introduced recently ~\cite{M00}.  It is now possible to follow
the direction corresponding to the negative eigenvalue exactly,
ensuring a full convergence onto the saddle point.

The configuration is first pushed along a random direction until a
negative eigenvalue appears.  At each step along this trajectory, its
total energy is relaxed in the hyperplane perpendicular to it.  This
ensure that the total energy and forces remain under control as the
configuration leaves the harmonic well.

Once the lowest eigenvalue passes a threshold (set here at 10$^{-3}$),
we start the convergence to the saddle point by pushing the
configuration along the eigenvector corresponding to this lowest
eigenvalue, while minimizing the forces in all other directions.
Unless the lowest eigenvalue turns positive, this procedure is
guaranteed to converge onto a first-order saddle point, where forces
on the configuration are zero.  If the lowest eigenvalue changes sign,
the iteration procedure is stopped and a new event is started.

Because ART is designed to work for systems with thousands of degrees
of freedom, it is not appropriate to perform a direct diagonalization
of the Hessian to extract eigenvalues and eigenvectors.  We use
instead the Lanczos algorithm~\cite{Lanc1,Lanc2}.  This algorithm
works by iteratively projecting a vector on the Hamiltonian,
extracting preferably the lowest eigenvalues and corresponding
eigenvectors.  In our case, 15 to 30 force evaluations are sufficient
to extract the very lowest eigenvalue and its eigenvector, requiring
the diagonalization of a trigonal matrix of the same dimensions.  As a
bonus, this iterative scheme can use the direction of the previous
step as a seed, ensuring a better convergence.

The second step of the algorithm, the relaxation, is
straightforward and can be achieved with any standard minimization
technique.  We use the conjugate gradient method (CG) ~\cite{CG}.

Since moves are defined directly in the 3N-dimensional
configurational space, ART is not sensitive to the constraints of
real space algorithms: a complex collective motion, requiring the
displacement of hundreds of atoms, is as easily produced as a
one-atom jump; and a high energy barrier does not require more
efforts to cross than a thermal one.  Such versatility is
particularly important in disordered and complex materials where
events can involve collective rearrangements that are hard to
foresee.

In this work, the energy landscape is described by the Lennard-Jones
potential:
\begin{equation} E = 4\epsilon
\sum_{i<j}\left[\left({\frac{\sigma}{{r}_{ij}}}\right)^{12}
-\left({\frac{\sigma}{{r}_{ij}}}\right)^{6}\right],
\end{equation}
where $\epsilon$ is the pair well depth and 2$^{1/6}\sigma$ is the
equilibrium pair separation.  The energy and distance are described
below in units of $\epsilon$ and $\sigma$, respectively.

As mentioned above, we compare our results to those obtained by
the Doye and Wales (DW) version of ART. DW propose a systematic
technique for exploring the surface energy landscape. Transition
states are found by using the eigenvector-following method
~\cite{eig1,eig2,eig3}, in which the energy is maximized along
one direction and simultaneously minimized in all the others. In
this approach, the Hessian matrix is diagonalized at the local
minimum and each of its eigenvectors are followed in turn in both
directions away from this minimum.  Although there is no
information regarding the position of saddle points in the
Hessian at the local minimum,  the eigenvalue of the eigenvectors
followed from this point often moves down and, a some point, might
becomes negative. From then on, the procedure is similar to that
described above.

The main advantage of this algorithm is that it provides a
systematic way of exploring the local energy landscape, moreover,
we can expect that its biases will be different from those of ART
nouveau described above.  It suffers some limitations, however.
First, the number of trial direction is finite, leading to a
maximum of 6N saddle points.  As we will see below, even in small
clusters, this is not enough to sample all saddle points. Second,
the method requires the full diagonalization of the Hessian
matrix repeatedly, at least in its first stage, making it an
order $N^3$ technique, too costly for problems of more than a few
hundred degrees of freedom.

\section{Simulations}

Simulations are done for the 13-atoms, 38-atoms and 55-atoms
Lennard-Jones clusters, using both ART and DW. In all cases, we
start from a relatively well-relaxed generic configuration, i.e.,
one which does not have special symmetries, and explore the
energy landscape around this minimum. The goal here is not to
recreate the full connectivity tree  --this was done already by
Doye et {\it al} \cite{funnel}--- but to study the biases of the
methods in finding events.

\subsection{A comparison between ART and DW}

For DW, the number of search directions is limited to $6N$.  A number
of these directions do not converge to a saddle point or lead to
degenerate activated points, producing of order $N$ structurally
different saddles.  Contrary to DW, ART can generate an infinite
number of initial search directions.  In the first part of the
simulation, we limit ourselves to sets of 3000 trial events, starting
from the same initial minimum,  for each cluster size.

After eliminating degenerate saddle configurations, ART is found to
have generated three to four times more events than DW for these clusters.
This ratio is, of course, related to the number of trial directions
used in the ART simulation.  As discussed below, for example, a 20 000
trial run on the 13-atom cluster can find all first-order paths
leaving a given local minimum.

Results for these runs, including statistics on the saddle points and
new configurations are reported in Table \ref{tab:tableau}.  In all
cases, permutational isomers are eliminated and only structurally
different configurations are counted.

\begin{table}
\caption{Number of structurally different saddle-points and new minima
(i.e. after eliminating configurational isomers) as a function of the
number of trial directions for the 13-atoms, the 38-atoms and the
55-atoms LJ clusters. For DW, all 6N possible directions are tried. For
ART, results presented here are based on 3000-event runs. Ranges of
activation and asymmetry energies (minimum and maximum) are also given
in unit of $\epsilon$.}
\label{tab:tableau}
\begin{tabular}{l|c|c|c|c}
Methods  &          &  DW         &  ART       & Common \\
\hline
13-atoms & Saddles & {\bf 17}    &  {\bf 72}  &   {\bf 13}\\
          & New minima    & {\bf 13}    &  {\bf 44}  &   {\bf 13} \\
          & Activation  & {\bf 0.54-3.62} & {\bf 0.54-3.68} \\
          & Asymmetry   & {\bf (-0.05)-2.75} & {\bf (-0.05)-3.57} \\
\hline
38-atoms & Saddles & {\bf 28}    &  {\bf 109}  &   {\bf 23}\\
          & New minima    & {\bf 21}    &  {\bf 73}  &   {\bf 21} \\
          & Activation & {\bf 0.26-3.2} & {\bf 0.14-5.66} \\
          & Asymmetry   & {\bf (-0.88)-2.64} & {\bf (-1.64)-3.1} \\
\hline
55-atoms & Saddles & {\bf 42}    &  {\bf 151}  &   {\bf 35}\\
          & New minima    & {\bf 29}    &  {\bf 89}  &   {\bf 29} \\
          & Activation  & {\bf 0.65-5.98} & {\bf 0.65-9.34} \\
          & Asymmetry   & {\bf (-1.84)-3.98}  & {\bf (-1.84)-8.22} \\

\end{tabular}
\end{table}

The sets of saddle points and minima obtained by ART and DW are
obviously not independent and we find, as one would expect, a
significant overlap between them: about 80 \% of DW events are
also found by the 3000-attempt run of ART (while about 20 \% of
ART events are also found by DW).

We can do a similar analysis for the new minima found by the two
methods.  Table \ref{tab:tableau} shows that in many cases, a
number of different saddle points lead to the same final minimum.
Moreover, this degeneracy seems to increase with the size of the
cluster. (The existence of multiple paths connecting two
minima is much more common for open systems, such as clusters,
than for bulk materials.  In the later case, the constraints of
volume and continuity make it much more difficult to find many
paths connecting two events.)  For all cluster sizes, the
3000-trial event ART simulations finds about 3 times more minima
than the DW method,  a ratio similar to the saddle configurations.

\subsection{Ergodicity and reversibility}

\begin{figure}
\epsfxsize=8cm
\epsfbox{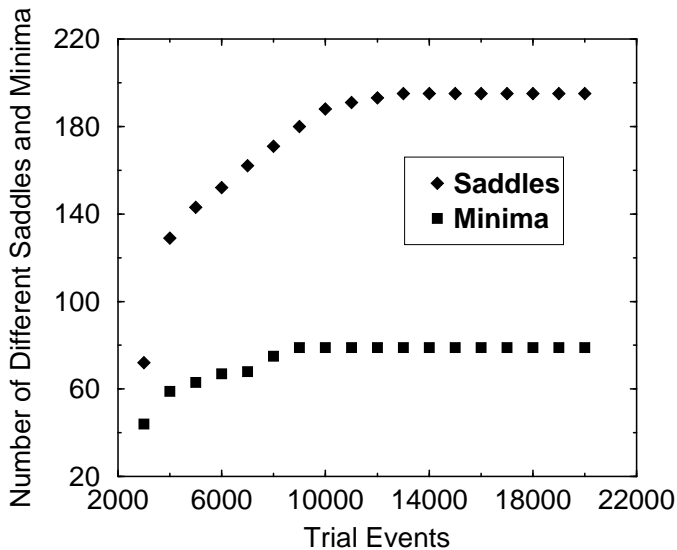}
\caption{Number of different saddle-points and minima as a function of
the number of trial events for a 20 000-attempt ART run the 13-atoms
LJ cluster.}
\label{fig:growth}
\end{figure}

It is possible to examine the question of the ergodicity of ART by
extending the simulations described above.  In Figure
\ref{fig:growth}, we trace the number of different saddle points and
minima as a function of trial events in a 20 000-attempt run, using a
single initial minimum as in the previous simulation.  The sampling of
minima seems to be complete after about 9000 events and we identify a
total of 79 different minima, including all those found using the DW
method.  It takes about 50 \% more events to generate all 195 saddle
points that can be found with ART. Here again, all saddle points
generated with DW are found by ART.
Although this does not show formally the completeness of the sets
found, the previous simulation, as well as its comparison with a
different technique, provides a fairly solid base for claiming
that ART is ergodic.

Dynamically, events should also be fully reversible.  We check this in
a two-step simulation.  First, we check the reversibility of the paths
from the saddle point.  After converging to a first order saddle
point, we pull the configuration back very slightly and apply the
relaxation routine from there.  For the 3000 ART-generated events on
the 13-atom LJ cluster, all but 10 (0.33\%) events relax to the
original minimum and these last 10 events converge to a different but
very close local minimum.  Saddle points found by ART are
therefore really delineating the boundary of the energy basin around
the initial minimum.

Clearly, paths are fully reversible
from their activated point.

The full trajectory must also be reversible; the initial minimum,
saddle, final minimum sequence should also be found in the
opposite order. For each LJ cluster, we select 20 new local
minima, reached in a one-step event from the initial minimum and
check that, indeed, the configuration can alway come back to that
state, and that the same saddle points are found in both
directions.  This important result allows us to conclude that the
whole pathway between local minima passing through first order
saddle-points is reversible.

\subsection{Stability under change of parameters}

\begin{figure}
\epsfxsize=8cm
\epsfbox{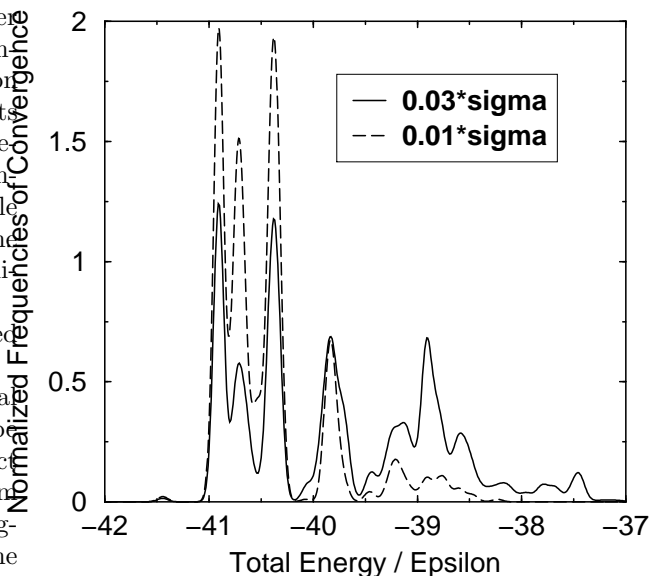}
\caption{Normalized frequencies of occurrence in the case of 13-atoms LJ
clusters for two step values, 0.01 and 0.03 $\sigma$.}
\label{1313}
\end{figure}

In the search for saddle points, a number of factors can
influence the selection of the activation paths. In particular,
it is important to verify that the step size along the direction
of the lowest eigenvalue, in the convergence to the saddle point,
does not result in some events missing. To examine this issue, we
perform the same simulation as that described above for the
13-atom LJ cluster with two different step values: 0.01 and 0.03
$\sigma$.  The distribution of energy barriers for all converged
saddle points is given in Fig. \ref{1313}.  All saddle points
found by the 0.01 $\sigma$ step are also found by the larger
step. Moreover, the 0.03 $\sigma$ seems to better converge to
higher energy barriers.  The total number of saddle-points reached
by using 0.03 $\sigma$ step value is also four times greater than that
found by 0.01 $\sigma$.  This sheds some light on the workings of the
algorithm. As we first start to follow the eigendirection
corresponding to the lowest eigenvalue, the configuration is in a
very shallow valley and too much relaxation perpendicular to this
valley can easily make it vanish. With a larger step moving away
from the minimum, the configuration reaches a deep valley faster,
increasing strongly the rate of convergence. It is clear, here,
that this parameter can be adjusted to optimize the rate of
convergence without fear of loosing particular saddle points.

\subsection{Biases in searching for saddle points}

We now address the question of biases in the search for saddle
points and new minima.  The range of activation and asymmetry
energies is quite wide for the events generated on the three
clusters, as can be seen in Table \ref{tab:tableau}.  As
expected, this range increases with the size of the system. This
behaviour is typical of clusters. For bulk system, the
distribution of relevant activation events is generally
independent of the size and is usually bounded by a small
multiple of the binding energy between two atoms.  The higher
bound for the distribution of activation energy reflects the
overall properties of the systems studied and should be rather
independent of the starting configuration; the lower bound,
however, is not; it represents a direct measure of the stability
of the initial metastable state.

\begin{figure}
\caption{(a) The distribution of saddle points and minima as a
function of energy/$\epsilon$ for the 55-atom clusters as found in a
3000-trial event ART run leading to 1856 completed events.  (b)
Normalized frequency of occurrent of saddle points as a function
of energy for the ART and DW runs for the same cluster.  The DW
run generated 109 successful events.  (c) Same as (b), but for
minima.  In both (b) and (c),  the solid curve represents ART and
the dashed one DW.} \label{fig:distr} \epsfxsize=8cm
\epsfbox{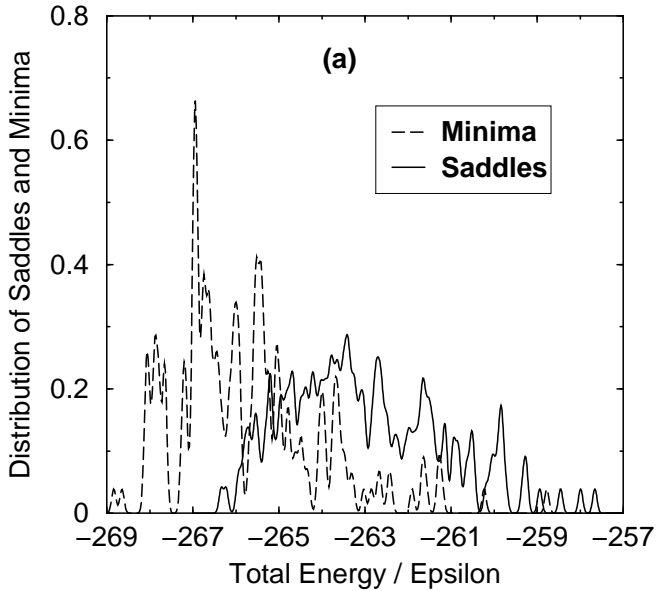}
\end{figure}

\begin{figure}
\epsfxsize=8cm
\epsfbox{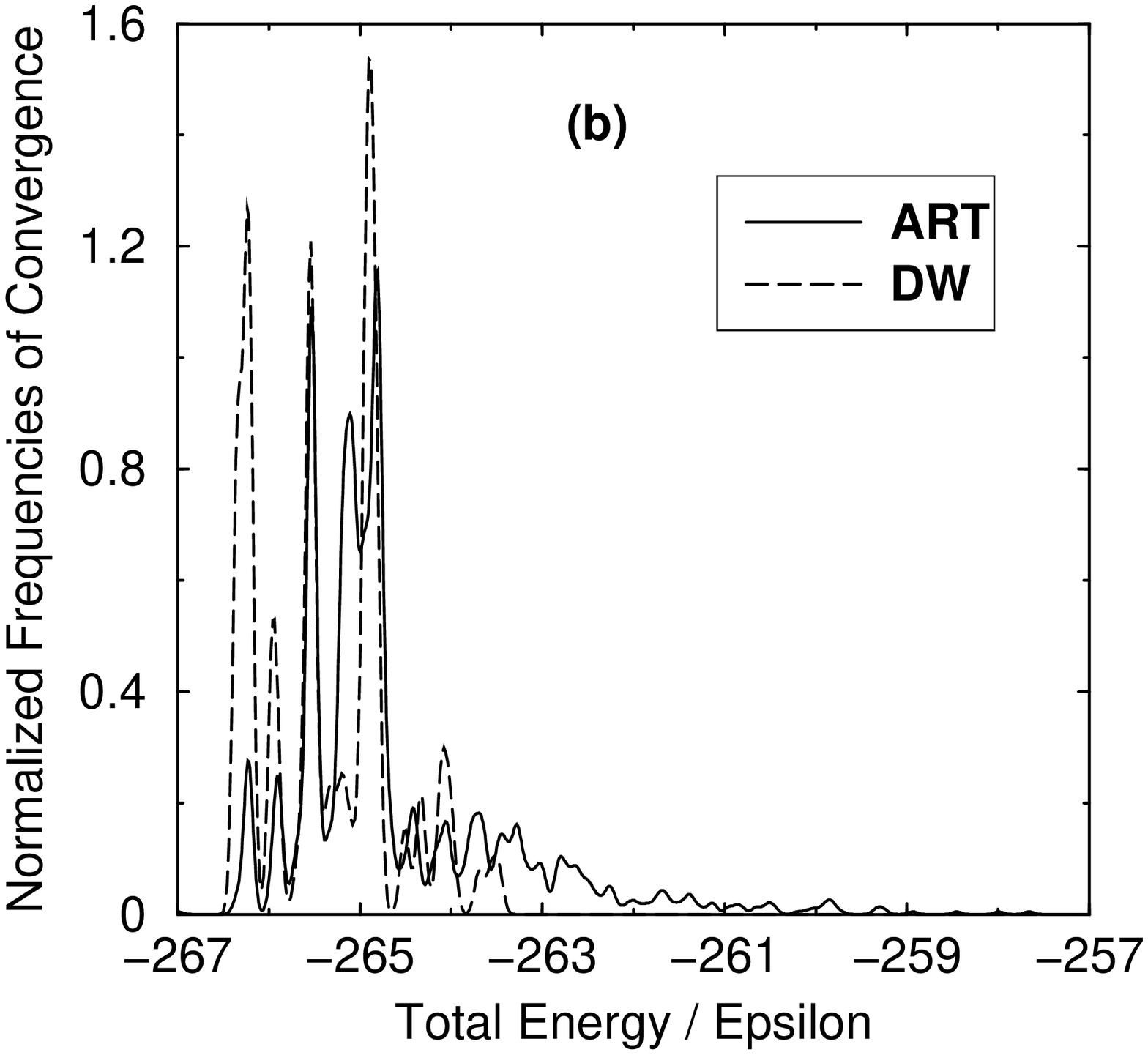}
\end{figure}

\begin{figure}
\epsfxsize=8cm
\epsfbox{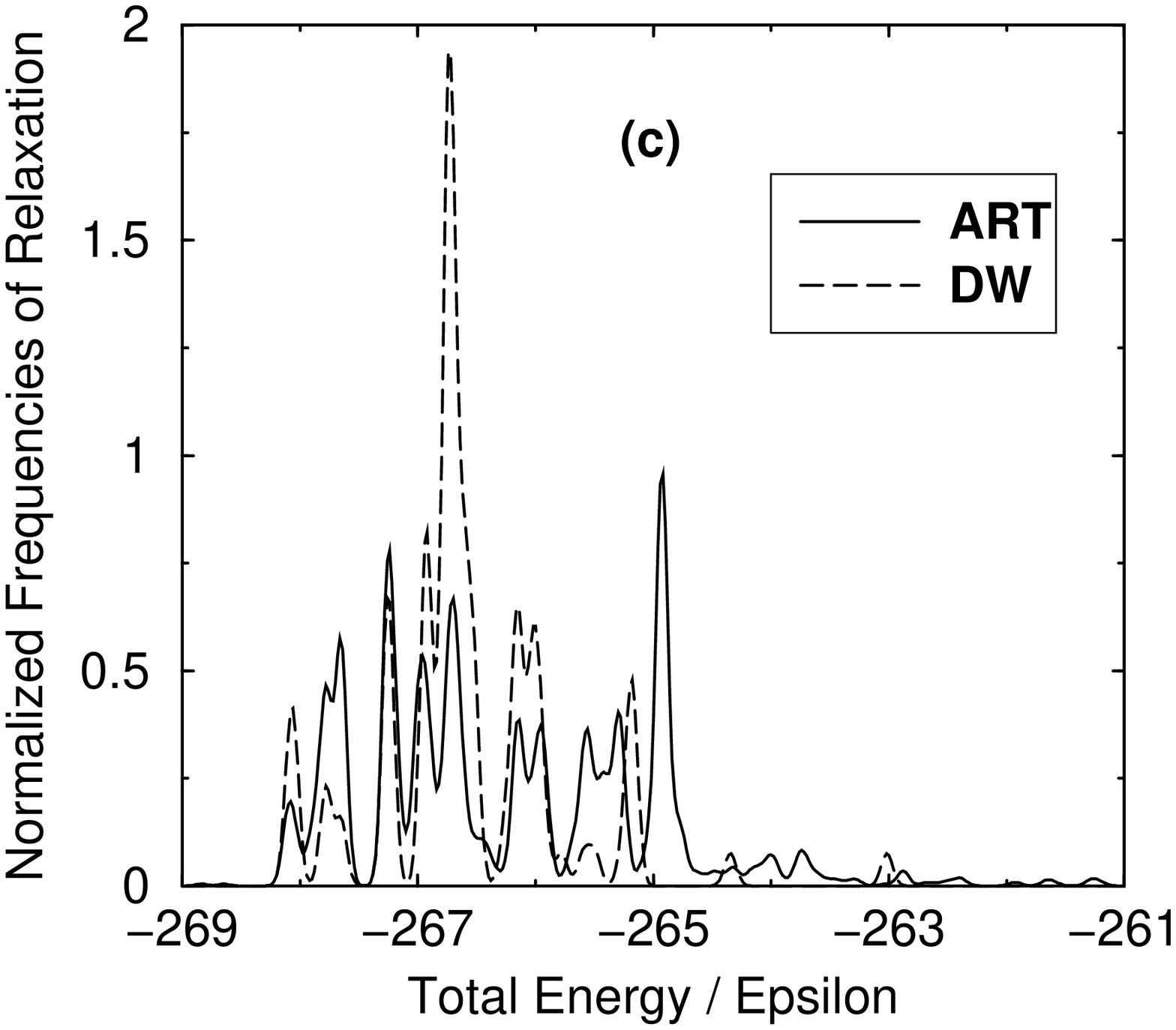}
\end{figure}

Figure \ref{fig:distr} gives two representations of the
distribution of activation and asymmetry energies for the 55-atom
clusters. In Fig. \ref{fig:distr}(a), we give the distribution of
energies for the 151 different saddles and 89 different minima
identified in the 3000-trial event run. The distribution of
saddle energies is rather smooth over its whole range while that
of the asymmetry energies presents a few peaks.

To study the biases of ART, we also plot in Fig.  \ref{fig:distr}(b)
and (c) the normalized distribution of activation and asymmetry
energies for {\it all} 1856 completed ART events and the corresponding
109 successful DW events (although there are a total of 330 trial
directions for DW, the method shows a poor success rate when
eigendirections corresponding to very high eigenvalues are selected.)
The distribution for DW is therefore almost discretized and peaks
correspond to single saddle points being visited many times.  The ART
distribution is almost continuous and we have binned the events as a
function of energy.

The structure of these distributions is quite different from that
of Fig. \ref{fig:distr}(a). In particular, there are
strongly peaked at low energies indicating that each method
seems to enhance significantly a few paths over the other ones.
The biases are not exactly the same,  however, although there is
considerable overlap for the two methods. This suggests that the
topography of the energy landscape around the minimum is
reflected, at least in part, in the choice of events.

\begin{figure}
\epsfxsize=8cm
\epsfbox{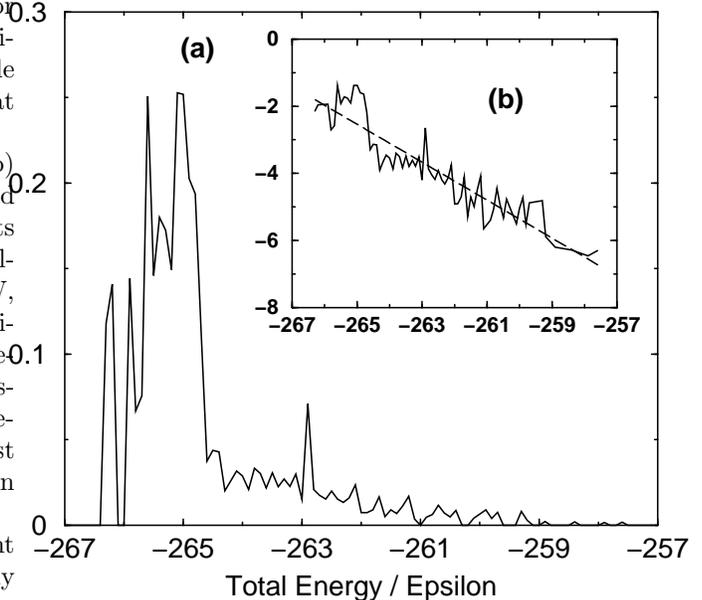} \vspace{0.5cm}

\caption{ (a) Ratio of the distribution of barriers for all saddle points
generated in a 3000-attemp ART run (Fig.  \ref{fig:distr}(a)) over
that of the different saddle points existing around a minimum (Fig.
\ref{fig:distr}(b).   The same distribution is plotted in inset (b)
with a log-normal scale. The dashed line is a fit with slope -0.57.}
\label{biase}
\end{figure}

\begin{figure}
\epsfxsize=8cm
\epsfbox{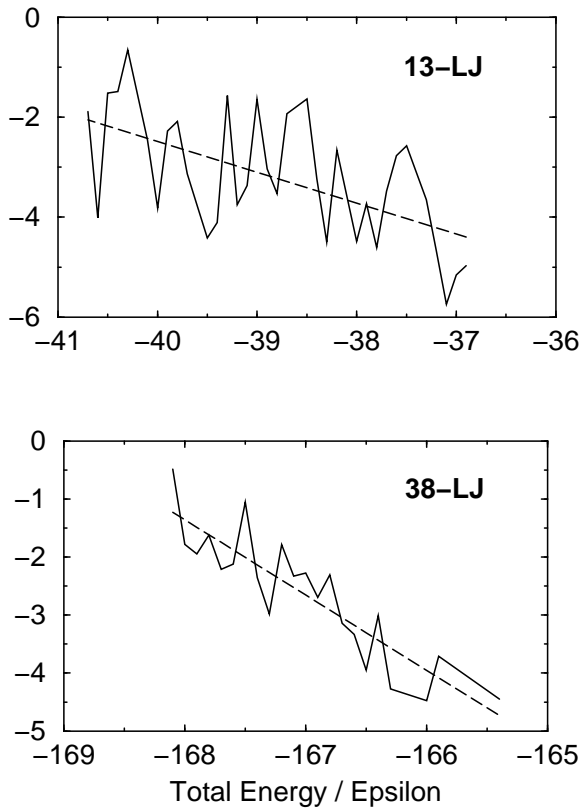}
\vspace{1cm}
\caption{Same as previous figure for the 13-atom and the 38-atom LJ
clusters. The slope of the fitted line is $-0.62$ and $-1.3$,
respectively.}
\label{setbiase}
\end{figure}

One approach to identify the overall biases of ART is to plot the
ratio of the distribution of barriers for all saddle points
generated (frequency of convergence) over that of the different
saddle points existing around our minimum.  This ratio, plotted
in Fig. \ref{biase}, provides a first indication on how ART selects
saddle points. In Fig. \ref{biase}(a), the
distribution is dominated by a strong peak at low activation
energy, followied by a fairly extended tail. Surprisingly, most
of the small structure that can be found in \ref{fig:distr}(a)
has been eliminated in the ratio, indicating that there is fairly
little bias towards a few events in particular but that the
selection of an event over another one is mostly a matter of
energy. This can be seen more clearly in the log-normal
distribution plotted in Fig. \ref{biase} (b). The result is
remarkable: the distribution is well fitted by an exponential
function $p(E) \propto \exp( -0.57 * E)$ and the sampling follows
a Boltzmann distribution!  The same behavior can be observed for
the 13-atom and the 38-atom clusters, as shown in Fig.
\ref{setbiase}.

\section{CONCLUSION}

The configurational energy landscape of Lennard-Jones clusters was
explored by using the activation relaxation technique (ART).
Comparing ART with a different sampling method proposed by Doye and
Wales, it does not seem that ART missed any class of first-order
activated paths.  Based on this and extensive simulations, we conclude
that ART can find all first-order saddle points and new minima around
a given minimum, indicating that it is ergodic.  The trajectories it
defines, initial minimum--saddle point--final minimum, are also
reversible, indicating that the trajectories found are real activated
paths.

We find, finally, that ART samples the surface energy landscape
according to a Boltzmann distribution, i.e., that the probability of
finding a given saddle point is proportional to $\exp(E_{barrier})/E_0$.
The original of this bias is not yet understood but this opens the
door to real-time activated dynamics in complex system, a very
exciting prospect.

\section{Acknowledgment}

We acknowledge useful discussions with G. Barkema as well as partial
support from NSF under grant number DMR-9805848.

\bibliographystyle{prsty}

\end{document}